\documentstyle[amssymb,floats,aps,prb,epsf]{revtex}
\epsfclipon 
\begin{document}
\twocolumn[\hsize\textwidth\columnwidth\hsize\csname
@twocolumnfalse\endcsname 

\draft

\title{Kinetic energy driven superconductivity in doped cuprates}
\author{Shiping Feng}
\address{Department of Physics and Key Laboratory of Beam
Technology and Material Modification, Beijing Normal University,
Beijing 100875, China}
\maketitle
\centerline{Received 30 June 2003}
\begin{abstract}
Within the $t$-$J$ model, the mechanism of superconductivity in
doped cuprates is studied based on the partial charge-spin
separation fermion-spin theory. It is shown that dressed holons
interact occurring directly through the kinetic energy by
exchanging dressed spinon excitations, leading to a net attractive
force between dressed holons, then the electron Cooper pairs
originating from the dressed holon pairing state are due to the
charge-spin recombination, and their condensation reveals the
superconducting ground state. The electron superconducting
transition temperature is determined by the dressed holon pair
transition temperature, and is proportional to the concentration
of doped holes in the underdoped regime. With the common form of
the electron Cooper pair, we also show that there is a coexistence
of the electron Cooper pair and antiferromagnetic short-range
correlation, and hence the antiferromagnetic short-range
fluctuation can persist into the superconducting state. Our results
are qualitatively consistent with experiments.
\end{abstract}
]
\bigskip

\narrowtext

Since the discovery of high-temperature superconductivity (HTSC)
in doped cuprates, much effort has concentrated on the
superconducting (SC) mechanism \cite{anderson1}. Much experimental
evidence, including the factor of $2e$ occurring in the flux
quantum and in the Josephson effect, as well as the electrodynamic
and thermodynamic properties, supports the pairing theory
\cite{tsuei}. The single common feature of cuprate superconductors
is the presence of the two-dimensional (2D) CuO$_{2}$ plane
\cite{kastner}, then it is believed that the relatively high SC
transition temperature $T_{c}$ is closely related to doped
CuO$_{2}$ planes. The undoped state of cuprate superconductors is
a Mott insulator with antiferromagnetic (AF) long-range order
(AFLRO), then changing the carrier concentration by ionic
substitution or increasing the oxygen content turns these compounds
into the SC state leaving the AF short-range correlation still
intact \cite{kastner}. Moreover, the superfluid density in the
underdoped regime vanishes more or less linearly with doping, and
the SC transition temperature $T_{c}$ is proportional to a positive
power of the concentration of doped holes $\delta$
($T_{c}\propto \delta$ in doped CuO$_{2}$) \cite{uemura,tallon}.
Therefore there is a general consensus that the HTSC to holes
interaction via a magnetic medium, and short-range AF
correlation coexists with the SC state.

In conventional superconductors, the electrons interact by
exchanging phonons. Since this interaction leads to a net
attractive force between electrons, the system can lower its
potential energy by forming electron Cooper pairs \cite{bcs}. These
electron Cooper pairs condense into a coherent macroscopic quantum
state and can move freely without resistance. As a result, pairing
in the conventional superconductors is always related to an
increase in kinetic energy which is overcompensated for by the
lowering of the potential energy \cite{bcs}. On the contrary, it
has been argued that the SC transition in doped cuprates is
determined by the need to reduce the frustrated kinetic energy
\cite{anderson2}, where the driving attractive force between holes
may be attributed to the fact that by sharing a common link two
holes minimize the loss of the energy related to breaking AF
links, and is therefore mediated by the exchange of spin
excitations \cite{dagotto}. Within the 2D $t$-$J$ model, robust
indications of superconductivity have been found by using numerical
techniques \cite{sorella}. Moreover, by solving a model for alkali
doped fullerenes within dynamical mean-field (MF) theory, it has
been argued \cite{capone} recently that the strong electron
correlation does not suppress superconductivity, but rather seems
to favor it because the main ingredient was identified as a
pairing mechanism not involving the charge density operator, but
other internal degrees of freedom, like the spin, unveiling a kind
of the charge-spin separation. These scenarios are consistent with
recent optical experiments \cite{molegraaf}. The normal-state above
$T_{c}$ exhibits a number of anomalous properties which are due to
charge-spin separation \cite{anderson1,anderson2}, while the SC
state is characterized by charge-spin recombination.

Recently, we \cite{feng1} have developed a partial charge-spin
separation fermion-spin theory to study the physical properties of
doped cuprates, where the electron operator is decoupled as the
gauge invariant dressed holon and spinon. Based on this theory, we
have discussed the unusual normal-state properties of the
underdoped cuprates, and the results are good consistent with the
experiments. It is shown that the charge transport is mainly
governed by the scattering from the dressed holons due to the
dressed spinon fluctuation, while the scattering from the dressed
spinons due to the dressed holon fluctuation dominates the spin
response. In this paper, we apply this approach to discuss the
mechanism of HTSC. Within the $t$-$J$ model, we show that dressed
holons interact occurring directly through the kinetic energy by
exchanging dressed spinon excitations, leading to a net attractive
force between dressed holons, then the electron Cooper pairs
originating from the dressed holon pairing state are due to the
charge-spin recombination, and their condensation reveals the SC
ground state. The SC transition temperature $T_{c}$ is identical
to the dressed holon pair transition temperature, and is
proportional to the concentration of doped holes in the underdoped
regime. With the common form of the electron Cooper pair, we also
show that there is a coexistence of the electron Cooper pair and AF
short-range correlation, and hence the AF short-range fluctuation
can persist into the SC state.

We start from the 2D $t$-$J$ model,
\begin{equation}
H=-t\sum_{i\hat{\eta}\sigma}C^{\dagger}_{i\sigma}
C_{i+\hat{\eta}\sigma}+\mu \sum_{i\sigma}
C^{\dagger}_{i\sigma}C_{i\sigma}+J\sum_{i\hat{\eta}}{\bf S}_{i}
\cdot {\bf S}_{i+\hat{\eta}},
\end{equation}
acting on the Hilbert space with no doubly occupied site, i.e.,
$\sum_{\sigma}C^{\dagger}_{i\sigma}C_{i\sigma}\leq 1$, where
$\hat{\eta}=\pm\hat{x},\pm\hat{y}$, $C^{\dagger}_{i\sigma}$
($C_{i\sigma}$) is the electron creation (annihilation) operator,
${\bf S}_{i}=C^{\dagger}_{i}{\vec\sigma}C_{i}/2$ is spin operator
with ${\vec\sigma}=(\sigma_{x},\sigma_{y},\sigma_{z})$ as Pauli
matrices, and $\mu$ is the chemical potential. In the $t$-$J$ model
(1), the strong electron correlation manifests itself by a single
occupancy local constraint, and therefore the crucial requirement
is to impose this local constraint. It has been shown that this
constraint can be treated properly in analytical calculations
within the partial charge-spin separation fermion-spin theory
\cite{feng1},
\begin{eqnarray}
C_{i\uparrow}=h^{\dagger}_{i\uparrow}S^{-}_{i},~~~~
C_{i\downarrow}=h^{\dagger}_{i\downarrow}S^{+}_{i},
\end{eqnarray}
with the {\it spinful fermion} operator $h_{i\sigma}=
e^{-i\Phi_{i\sigma}}h_{i}$ describes the charge degree of freedom
together with the phase part of the spin degree of freedom ({\it
dressed holon}), while the spin operator $S_{i}$ describes the
amplitude part of the spin degree of freedom ({\it dressed
spinon}), then the electron on-site local constraint for the
single occupancy, $\sum_{\sigma}C^{\dagger}_{i\sigma}
C_{i\sigma}=S^{+}_{i}h_{i\uparrow}h^{\dagger}_{i\uparrow}S^{-}_{i}
+S^{-}_{i}h_{i\downarrow}h^{\dagger}_{i\downarrow}S^{+}_{i}=h_{i}
h^{\dagger}_{i}(S^{+}_{i}S^{-}_{i}+S^{-}_{i}S^{+}_{i})=1-
h^{\dagger}_{i}h_{i}\leq 1$, is satisfied in analytical
calculations, and the double {\it spinful fermion} occupancies,
$h^{\dagger}_{i\sigma}h^{\dagger}_{i-\sigma}=e^{i\Phi_{i\sigma}}
h^{\dagger}_{i}h^{\dagger}_{i}e^{i\Phi_{i-\sigma}}=0$,
$h_{i\sigma}h_{i-\sigma}=e^{-i\Phi_{i\sigma}}h_{i}h_{i}
e^{-i\Phi_{i-\sigma}}=0$, are ruled out automatically. As we have
shown in Ref. \cite{feng1}, the phase factor $\Phi_{i\sigma}$ is
separated from the {\it bare spinon} operator, therefore it also
is an operator, and describes a spinon cloud. It has been shown
that these dressed holons and spinons are gauge invariant, and in
this sense, they are real. This dressed holon $h_{i\sigma}$ is
a {\it spinless fermion} $h_{i}$ ({\it bare holon}) incorporating
the spinon cloud $e^{-i\Phi_{i\sigma}}$ (magnetic flux), and is
a magnetic dressing. In other words, the gauge-invariant dressed
holon carries some spinon messages, i.e., it shares some effects
of the spinon configuration rearrangements due to the presence of
the hole itself \cite{martins}. Although in common sense
$h_{i\sigma}$ is not a real {\it spinful fermion}, it behaves like
a {\it spinful fermion}. The spirit of the partial charge-spin
separation fermion-spin theory is that the electron operator can be
mapped as a product of the spin operator and spinful fermion
operator, this is very similar to those of bosonization in
one-dimensional interacting electron systems, where the electron
operators are mapped onto the boson (electron density)
representation, and then the recast Hamiltonian is exactly
solvable. In this partial charge-spin separation fermion-spin
representation, the low-energy behavior of the $t$-$J$ model (1)
can be expressed as \cite{feng1},
\begin{eqnarray}
H&=&-t\sum_{i\hat{\eta}}(h_{i\uparrow}S^{+}_{i}
h^{\dagger}_{i+\hat{\eta}\uparrow}S^{-}_{i+\hat{\eta}}+
h_{i\downarrow}S^{-}_{i}h^{\dagger}_{i+\hat{\eta}\downarrow}
S^{+}_{i+\hat{\eta}})\nonumber \\
&-&\mu\sum_{i\sigma}h^{\dagger}_{i\sigma}h_{i\sigma}+J_{{\rm eff}}
\sum_{i\hat{\eta}}{\bf S}_{i}\cdot {\bf S}_{i+\hat{\eta}},
\end{eqnarray}
with $J_{{\rm eff}}=(1-\delta)^{2}J$, and $\delta=\langle
h^{\dagger}_{i\sigma}h_{i\sigma}\rangle=\langle h^{\dagger}_{i}
h_{i}\rangle$ is the hole doping concentration. As a consequence,
the kinetic energy ($t$) term in the $t$-$J$ model has been
expressed as the dressed holon-spinon interaction, which dominates
the essential physics of doped cuprates, while the magnetic energy
($J$) term is only to form an adequate dressed spinon configuration
\cite{anderson2}. The SC state is characterized by electron Cooper
pairs, forming SC quasiparticles \cite{tsuei}, and the order
parameter for the electron Cooper pair can be expressed as,
\begin{eqnarray}
\Delta=\langle C^{\dagger}_{i\uparrow}C^{\dagger}_{j\downarrow}-
C^{\dagger}_{i\downarrow}C^{\dagger}_{j\uparrow}\rangle =\langle
h_{i\uparrow}h_{j\downarrow}S^{+}_{i}S^{-}_{j}-h_{i\downarrow}
h_{j\uparrow}S^{-}_{i}S^{+}_{j}\rangle .
\end{eqnarray}
In the doped regime without AFLRO, the dressed spinons form the
disordered spin liquid state, where the dressed spinon correlation
function $\langle S^{+}_{i}S^{-}_{j}\rangle=\langle S^{-}_{i}
S^{+}_{j}\rangle$, then the order parameter for the electron Cooper
pair in Eq. (4) can be written as
$\Delta=-\langle S^{+}_{i}S^{-}_{j} \rangle\Delta_{h}$, with the
dressed holon pairing order parameter $\Delta_{h}=\langle
h_{j\downarrow}h_{i\uparrow}-h_{j\uparrow}h_{i\downarrow}\rangle$.
In this case, the physical properties of the SC state are
essentially determined by the dressed holon pairing state. However,
in the extreme low doped regime with AFLRO, where the dressed
spinon correlation function
$\langle S^{+}_{i}S^{-}_{j}\rangle\neq\langle S^{-}_{i}S^{+}_{j}
\rangle$, the conduct is disrupted by AFLRO. Therefore in the
following discussions, we only focus on the case without AFLRO.

The quantum spin operators obey the Pauli spin algebra, and this
problem can be discussed in terms of the two-time spin Green's
function within the Tyablikov scheme \cite{tyablikov}. In this
case, we define the dressed holon diagonal and off-diagonal
Green's functions as,
\begin{mathletters}
\begin{eqnarray}
g(i-j,t-t')&=&-i\theta(t-t')\langle [h_{i\sigma}(t),
h^{\dagger}_{j\sigma}(t')]\rangle \nonumber \\
&=&\langle\langle h_{i\sigma}(t);h^{\dagger}_{j\sigma}(t') \rangle
\rangle ,\\
\Im (i-j,t-t')&=&-i\theta(t-t')\langle [h_{i\downarrow}(t),
h_{j\uparrow}(t')]\rangle \nonumber \\
&=&\langle\langle h_{i\downarrow}(t);h_{j\uparrow}(t') \rangle
\rangle ,\\
\Im^{\dagger} (i-j,t-t')&=&-i\theta(t-t')\langle
[h^{\dagger}_{i\uparrow}(t),h^{\dagger}_{j\downarrow}(t')]\rangle
\nonumber \\
&=&\langle\langle h^{\dagger}_{i\uparrow}(t);
h^{\dagger}_{j\downarrow}(t')\rangle\rangle ,
\end{eqnarray}
\end{mathletters}
and the dressed spinon Green's functions as,
\begin{mathletters}
\begin{eqnarray}
D(i-j,t-t')&=&-i\theta(t-t')\langle [S^{+}_{i}(t),S^{-}_{j}(t')]
\rangle \nonumber \\
&=&\langle\langle S^{+}_{i}(t);S^{-}_{j}(t')\rangle\rangle, \\
D_{z}(i-j,t-t')&=&-i\theta(t-t')\langle [S^{z}_{i}(t),S^{z}_{j}
(t')]\rangle \nonumber \\
&=&\langle\langle S^{z}_{i}(t);S^{z}_{j}(t')\rangle\rangle,
\end{eqnarray}
\end{mathletters}
respectively, where $\langle \ldots \rangle$ is an average over
the ensemble. In the MF level, the dressed spinon system is an
anisotropic away from the half-filling \cite{feng2}, therefore we
have defined the two dressed spinon Green's functions $D(i-j,t-t')$
and $D_{z}(i-j,t-t')$ to describe the dressed spinon propagations.
In the doped regime without AFLRO, i.e.,
$\langle S^{z}_{i}\rangle=0$, a MF theory of the $t$-$J$ model
based on the fermion-spin theory has been developed \cite{feng2}
within the Kondo-Yamaji decoupling scheme \cite{kondo}, which is a
stage one-step further than Tyablikov's decoupling scheme. In this
MF theory \cite{feng2}, the phase factor $e^{i\Phi_{i\sigma}}$
describing the phase part of the spin degree of freedom was not
considered. Following their discussions \cite{feng2}, we can obtain
the MF dressed holon and spinon Green's functions in the present
partial charge-spin separation fermion-spin theoretical framework
as,
\begin{mathletters}
\begin{eqnarray}
g^{(0)}(k)&=&{1\over i\omega_{n}-\xi_{{\bf k}}}, \\
D^{(0)}(p)&=&{B_{{\bf p}}\over (ip_{m})^{2}-\omega_{{\bf p}}^{2}}
\nonumber \\
&=&{1\over 2}\sum_{\nu=1,2}{B_{{\bf p}}\over\omega_{\nu}({\bf p})}
{1\over ip_{m}-\omega_{\nu}({\bf p})}, \\
D^{(0)}_{z}(p)&=&{B_{z}({\bf p})\over (ip_{m})^{2}-\omega_{z}
({\bf p})^{2}} \nonumber \\
&=&{1\over 2}\sum_{\nu=1,2}{B_{z}({\bf p})\over\omega_{z\nu}
({\bf p})}{1\over ip_{m}-\omega_{z\nu}({\bf p})},
\end{eqnarray}
\end{mathletters}
respectively, where the four-vector notation
$k=({\bf k},i\omega_{n})$, $p=({\bf p},ip_{m})$,
$B_{{\bf p}}=\lambda[2\chi^{z}(\epsilon\gamma_{{\bf p}}-1)+\chi
(\gamma_{{\bf p}}-\epsilon)]$, $B_{z}({\bf p})=\lambda\chi\epsilon
(\gamma_{{\bf p}}-1)$, $\lambda=2ZJ_{eff}$, $\gamma_{{\bf p}}=(1/Z)
\sum_{\hat{\eta}}e^{i{\bf p}\cdot\hat{\eta}}$,
$\epsilon=1+2t\phi/J_{{\rm eff}}$, $Z$ is the number of the nearest
neighbor sites, $\omega_{1}({\bf p})=\omega_{{\bf p}}$,
$\omega_{2}({\bf p})=-\omega_{{\bf p}}$,
$\omega_{z1}({\bf p})=\omega_{z}({\bf p})$,
$\omega_{z2}({\bf p})=-\omega_{z}({\bf p})$, and the MF dressed
holon and spinon excitation spectra are given by,
\begin{mathletters}
\begin{eqnarray}
\xi_{{\bf k}}&=&Zt\chi\gamma_{{\bf k}}-\mu, \\
\omega^{2}_{{\bf p}}&=&A_{1}\gamma^{2}_{{\bf p}}+A_{2}
\gamma_{{\bf p}}+A_{3}, \\
\omega^{2}_{z}({\bf p})&=&\epsilon\lambda^{2}(A_{z}-\alpha\chi
\gamma_{{\bf p}})(1-\gamma_{{\bf p}}),
\end{eqnarray}
\end{mathletters}
respectively, with $A_{1}=\alpha\epsilon\lambda^{2}(\epsilon
\chi^{z}+\chi/2)$, $A_{2}=-\epsilon\lambda^{2}[\alpha(\chi^{z}+
\epsilon\chi/2)+(\alpha C^{z}+(1-\alpha)/(4Z)-\alpha\epsilon
\chi/(2Z))+(\alpha C+(1-\alpha)/(2Z)-\alpha\chi^{z}/2)/2]$,
$A_{3}=\lambda^{2}[\alpha C^{z}+(1-\alpha)/(4Z)-\alpha\epsilon
\chi/(2Z)+\epsilon^{2}(\alpha C+(1-\alpha)/(2Z)-\alpha
\chi^{z}/2)/2]$, $A_{z}=\epsilon[\alpha C+(1-\alpha)/(2Z)]-
\alpha\chi/Z$, and the dressed holon's particle-hole parameters
$\phi=\langle h^{\dagger}_{i\sigma}h_{i+\hat{\eta}\sigma}\rangle$,
the dressed spinon correlation functions
$\chi=\langle S_{i}^{+}S_{i+\hat{\eta}}^{-}\rangle$,
$\chi^{z}=\langle S_{i}^{z}S_{i+\hat{\eta}}^{z}\rangle$,
$C=(1/Z^{2})\sum_{\hat{\eta},\hat{\eta'}}\langle
S_{i+\hat{\eta}}^{+}S_{i+\hat{\eta'}}^{-}\rangle$, and
$C^{z}=(1/Z^{2})\sum_{\hat{\eta},\hat{\eta'}}\langle
S_{i+\hat{\eta}}^{z}S_{i+\hat{\eta'}}^{z}\rangle$.
In order not to violate the sum rule of the correlation function
$\langle S^{+}_{i}S^{-}_{i}\rangle=1/2$ in the case without
AFLRO, the important decoupling parameter $\alpha$ has been
introduced in the above MF calculation \cite{feng2,kondo}, which
can be regarded as the vertex correction.

In the $t$-$J$ model (3), the dressed holon-spinon coupling
occurring in the kinetic energy term is quite strong. This
interaction (kinetic energy) can induce the dressed holon pairing
state (then the electron pairing state and superconductivity) by
exchanging dressed spinon excitations in a higher power of the hole
doping concentration $\delta$. For a discussion of this problem, we
follow Eliashberg's strong coupling theory \cite{eliashberg},
and obtain the self-consistent equations in terms of the equation
of motion method \cite{zubarev,feng1} which is satisfied by the
full dressed holon diagonal and off-diagonal Green's functions as,
\begin{mathletters}
\begin{eqnarray}
g(k)&=&g^{(0)}(k) \nonumber \\
&+&g^{(0)}(k)[\Sigma^{(h)}_{1}(k)g(k)-
\Sigma^{(h)}_{2}(-k)\Im^{\dagger}(k)], \\
\Im^{\dagger}(k)&=&g^{(0)}(-k)\nonumber \\
&\times& [\Sigma^{(h)}_{1}(-k)\Im^{\dagger}(-k)
+\Sigma^{(h)}_{2}(-k)g(k)],
\end{eqnarray}
\end{mathletters}
with the self-energies are evaluated as,
\begin{mathletters}
\begin{eqnarray}
\Sigma^{(h)}_{1}(k)&=&(Zt)^{2}{1\over N^{2}}\sum_{{\bf p,p'}}
\gamma^{2}_{{\bf p+p'+k}}{1\over \beta} \nonumber \\
&\times& \sum_{ip_{m}}g(p+k){1\over\beta}\sum_{ip'_{m}}D(p')
D(p'+p), \\
\Sigma^{(h)}_{2}(k)&=&(Zt)^{2}{1\over N^{2}}\sum_{{\bf p,p'}}
\gamma^{2}_{{\bf p+p'+k}}{1\over \beta} \nonumber \\
&\times& \sum_{ip_{m}}\Im (-p-k){1\over\beta}\sum_{ip'_{m}}D(p')
D(p'+p).
\end{eqnarray}
\end{mathletters}
The pairing force and dressed holon gap function have been
incorporated into the self-energy $\Sigma^{(h)}_{2}(k)$, therefore
the self-energy $\Sigma^{(h)}_{2}(k)$ is called as the {\it
effective dressed holon gap function}. Moreover, this self-energy
$\Sigma^{(h)}_{2}(k)$ is an even function of $i\omega_{n}$, while
the other self-energy $\Sigma^{(h)}_{1}(k)$ is not. As we
\cite{feng1,feng5} have known from the discussion of the
normal-state properties, the self-energy $\Sigma^{(h)}_{1}(k)$
renormalizes the MF dressed holon spectrum, and therefore it
dominates the charge transport of the systems. As a qualitative
discussion, we neglect $\Sigma^{(h)}_{1}(k)$ in this paper, and
only study the static limit of the effective dressed holon gap
function, i.e., $\Sigma^{(h)}_{2}(k)=\bar{\Delta}_{h}({\bf k})$.
In this case, we obtain dressed holon diagonal and off-diagonal
Green's functions as,
\begin{mathletters}
\begin{eqnarray}
g(k)&=&{i\omega_{n}+\xi_{{\bf k}}\over (i\omega_{n})^{2}-
E^{2}_{{\bf k}}}\nonumber \\
&=&{1\over 2}\sum_{\nu=1,2}\left ( 1+{\xi_{{\bf k}}
\over E_{\nu}({\bf k})} \right ){1\over i\omega_{n}-
E_{\nu}({\bf k})} ,\\
\Im^{\dagger}(k)&=&-{\bar{\Delta}_{h}({\bf k})\over
(i\omega_{n})^{2}-E^{2}_{{\bf k}}} \nonumber \\
&=&-{1\over 2}\sum_{\nu=1,2}
{\bar{\Delta}_{h}({\bf k})\over E_{\nu}({\bf k})}{1\over
i\omega_{n}-E_{\nu}({\bf k})},
\end{eqnarray}
\end{mathletters}
with $E_{1}({\bf k})=E_{{\bf k}}$, $E_{2}({\bf k})=-E_{{\bf k}}$,
and the dressed holon quasiparticle spectrum $E_{{\bf k}}=
\sqrt{\xi^{2}_{{\bf k}}+\mid\bar{\Delta}_{h}({\bf k})\mid^{2}}$.
It has been shown \cite{shen1} from angle resolved photoemission
spectroscopy (ARPES) measurements that in real space the gap
function and pairing force have a range of one lattice spacing,
this indicates that the {\it effective dressed holon gap function}
can be expressed as $\bar{\Delta}_{h}({\bf k})=
\bar{\Delta}^{(a)}_{h}\gamma^{(a)}_{{\bf k}}$. On the other hand,
some experiments seem consistent with an s-wave pairing
\cite{chaudhari}, while other measurements gave evidence in favor
of d-wave pairing \cite{martindale,tsuei}, therefore in the
following discussions, we consider the cases of
$\bar{\Delta}^{(a)}_{h}=\bar{\Delta}^{(s)}_{h}$, and
$\gamma^{(a)}_{{\bf k}}=\gamma^{(s)}_{{\bf k}}=\gamma_{{\bf k}}=
({\rm cos}k_{x}+{\rm cos}k_{y})/2$, for s-wave pairing, and
$\bar{\Delta}^{(a)}_{h}=\bar{\Delta}^{(d)}_{h}$,
$\gamma^{(a)}_{{\bf k}}=\gamma^{(d)}_{{\bf k}}= ({\rm cos}k_{x}-
{\rm cos}k_{y})/2$, for d-wave pairing, respectively. Furthermore,
we limit ourselves to the MF level for the dressed spinon part,
since the normal-state charge transport obtained at this level can
well describe the experimental data \cite{feng1,feng5}. In this
case, we find from Eq. (10b) that the {\it effective dressed holon
gap parameter} satisfies the equation,
\begin{eqnarray}
1&=&-(Zt)^{2}{1\over N^{3}}\sum_{{\bf k,q,p}}\gamma^{2}_{{\bf k+q}}
\gamma^{(a)}_{{\bf k-p+q}}\gamma^{(a)}_{{\bf k}}
\sum_{\nu,\nu',\nu''}{1\over 2E_{\nu''}({\bf k})}\nonumber \\
&\times& {B_{{\bf q}}B_{{\bf p}}\over\omega_{\nu}({\bf q})
\omega_{\nu'}({\bf p})}{F_{\nu\nu'\nu''}({\bf k,q,p})\over
\omega_{\nu'}({\bf p})-\omega_{\nu}({\bf q})-E_{\nu''}({\bf k})},
\end{eqnarray}
where $F_{\nu\nu'\nu''}({\bf k,q,p})=n_{F}[E_{\nu''}({\bf k})]
(n_{B}[\omega_{\nu}({\bf q})]-n_{B}[\omega_{\nu'}({\bf p})])+
n_{B}[\omega_{\nu'}({\bf p})](1+n_{B}[\omega_{\nu}({\bf q})])$,
with $n_{B}[\omega_{\nu}({\bf p})]$ and $n_{F}[E_{\nu}({\bf k})]$
are the boson and fermion distribution functions, respectively.
This gap equation must be solved simultaneously with other
self-consistent equations,
\begin{mathletters}
\begin{eqnarray}
\phi &=& {1\over 2N}\sum_{{\bf k}}\gamma_{{\bf k}}
\left (1-{\xi_{{\bf k}}\over E_{{\bf k}}}{\rm th}
[{1\over 2}\beta E_{{\bf k}}]\right ),\\
\delta &=& {1\over 2N}\sum_{{\bf k}}\left (1-{\xi_{{\bf k}}
\over E_{{\bf k}}}{\rm th}[{1\over 2}\beta E_{{\bf k}}]
\right ),\\
\chi &=&{1\over N}\sum_{{\bf k}}\gamma_{{\bf k}}
{B_{{\bf k}}\over 2\omega_{{\bf k}}}{\rm coth}
[{1\over 2}\beta\omega_{{\bf k}}],\\
C &=&{1\over N}\sum_{{\bf k}}\gamma^{2}_{{\bf k}}
{B_{{\bf k}}\over 2\omega_{{\bf k}}}{\rm coth}
[{1\over 2}\beta\omega_{{\bf k}}],\\
{1\over 2} &=&{1\over N}\sum_{{\bf k}}{B_{{\bf k}}
\over 2\omega_{{\bf k}}}{\rm coth}
[{1\over 2}\beta\omega_{{\bf k}}],\\
\chi_{z}&=&{1\over N}\sum_{{\bf k}}\gamma_{{\bf k}}
{B_{z}({\bf k})\over 2\omega_{z}({\bf k})}{\rm coth}
[{1\over 2}\beta\omega_{z}({\bf k})],\\
C_{z}&=&{1\over N}\sum_{{\bf k}}\gamma^{2}_{{\bf k}}
{B_{z}({\bf k})\over 2\omega_{z}({\bf k})}{\rm coth}
[{1\over 2}\beta\omega_{z}({\bf k})],
\end{eqnarray}
\end{mathletters}
and therefore all the above order parameters, decoupling parameter
$\alpha$, and chemical potential $\mu$ are determined by the
self-consistent calculation \cite{feng2}, then the dressed holon
pair order parameter is obtained in terms of the off-diagonal
Green's function (11b) as,
\begin{eqnarray}
\Delta^{(a)}_{h}={2\over N}\sum_{{\bf k}}
[\gamma^{(a)}_{{\bf k}}]^{2}
{\bar{\Delta}^{(a)}_{h}\over E_{{\bf k}}}{\rm th}
[{1\over 2}\beta E_{{\bf k}}].
\end{eqnarray}

The dressed holon pairing state originating from the kinetic
energy term by exchanging dressed spinon excitations will also
lead to form the electron Cooper pairing state as mentioned in
Eq. (4). For a discussion of the physical properties of the SC
state, we need to calculate the electron off-diagonal Green's
function $\Gamma^{\dagger}(i-j,t-t')=\langle\langle
C^{\dagger}_{i\uparrow}(t);C^{\dagger}_{j\downarrow}(t')\rangle
\rangle$, which is a convolution of the dressed spinon Green's
function $D(p)$ and off-diagonal dressed holon Green's function
$\Im(k)$ in the framework of the partial charge-spin separation
fermion-spin theory, and can be expressed as,
\begin{eqnarray}
\Gamma^{\dagger}(k)={1\over N}\sum_{{\bf p}}{1\over \beta}
\sum_{ip_{m}}D(p)\Im(p-k).
\end{eqnarray}
This convolution of the dressed spinon Green's function and
off-diagonal dressed holon Green's function reflects the
charge-spin recombination \cite{anderson2}, and can be evaluated
in terms of the dressed spinon Green's function (7b) and
off-diagonal dressed holon Green's function (11b) as,
\begin{eqnarray}
\Gamma^{\dagger}(k)&=&-{1\over N}\sum_{{\bf p},\nu,\nu'}
{\bar{\Delta}^{(a)}_{h}({\bf p-k})\over 2E_{\nu'}({\bf p-k})}
\nonumber \\
&\times&{B_{{\bf p}}\over 2\omega_{\nu}({\bf p})}{n_{B}
[\omega_{\nu}({\bf p})]+n_{F}[E_{\nu'}({\bf p-k})]\over
i\omega_{n}-\omega_{\nu}({\bf p})+E_{\nu'}({\bf p-k})}.
\end{eqnarray}
With the help of this electron off-diagonal Green's function, we
can obtain the SC gap function as,
\begin{eqnarray}
\Delta^{(a)}({\bf k})&=&-{1\over N}\sum_{{\bf p}}
{\bar{\Delta}^{(a)}_{h}({\bf p-k})\over 2E_{{\bf p-k}}}{\rm th}
[{1\over 2}\beta E_{{\bf p-k}}]{B_{{\bf p}}\over 2\omega_{{\bf p}}}
\nonumber \\
&\times& {\rm coth}[{1\over 2}\beta\omega_{{\bf p}}],
\end{eqnarray}
which shows that the symmetry of the electron Cooper pair is
essentially determined by the symmetry of the dressed holon pair,
and therefore the SC gap function can be written as
$\Delta^{(a)}({\bf k})=\Delta^{(a)}\gamma^{(a)}_{{\bf k}}$,
with the SC gap parameter evaluated in terms of Eqs. (17) and
(14) as $\Delta^{(a)}=-\chi\Delta^{(a)}_{h}$. It has been shown
that the AF fluctuation is dominated by the scattering of dressed
spinons \cite{feng1,feng6}, while in the present case, this AF
fluctuation has been incorporated into the electron off-diagonal
Green's function (and hence the electron Cooper pair) in terms of
the dressed spinon Green's function. Since the form of the electron
Cooper pair (4) is common, and the off-diagonal electron Green's
function always is a convolution of the dressed spinon Green's
function and dressed holon off-diagonal Green's function in the
framework of the partial charge-spin separation fermion-spin
theory, there is a coexistence of the electron Cooper pair and
short-range AF correlation \cite{feng8}, and hence the short-range
AF fluctuation can persist into superconductivity,
which is consistent with the experiments \cite{kastner}. In Fig. 1,
we plot the dressed holon (a) and SC (b) gap parameters in the
s-wave symmetry (solid line) and d-wave symmetry (dashed line) as
a function of hole doping concentration $\delta$ at $T=0.005J$ and
$t/J=2.5$, where both dressed holon and SC gap parameters are
increased with increasing dopings. Although there is a coexistence
of the electron Cooper pair and short-range AF correlation, the
value of the SC gap parameter is still suppressed by this AF
fluctuation. Moreover, the range of
$\Delta^{(s)}_{h}$ ($\Delta^{(s)})$ in the s-wave symmetry is
initial from doping $\delta\approx 0$, while $\Delta^{(d)}_{h}$
($\Delta^{(d)})$ in the d-wave symmetry from $\delta\approx 0.05$
in the present case of $t/J=2.5$, and the value of
$\Delta^{(s)}_{h}$ ($\Delta^{(s)})$ is always larger than
$\Delta^{(d)}_{h}$ ($\Delta^{(d)})$ in the whole doped regime. The
present result in Eq. (17) also shows that the SC transition
temperature $T^{(a)}_{c}$ occurring in the case of $\Delta^{(a)}=0$
is identical to the dressed holon pair transition temperature
occurring in the case of $\bar{\Delta}^{(a)}_{h}=0$. In
correspondence with the SC gap parameter, the SC transition
temperature $T^{(a)}_{c}$ as a function of hole doping
concentration $\delta$ in the s-wave symmetry (solid line) and
d-wave symmetry (dashed line) for $t/J=2.5$ is plotted in Fig. 2
in comparison with the experimental result \cite{tallon} (inset).
Our results indicate that in the underdoped regime $T^{(a)}_{c}$ is
proportional to concentration of doped holes $\delta$, in
qualitative agreement with the experimental data
\cite{uemura,tallon}. These results can also be understood from the
properties of the dressed holon excitation spectrum $\xi_{{\bf k}}$
in Eq. (8a). At $T=T^{(a)}_{c}$, $\Delta^{(a)}=\Delta^{(a)}_{h}=0$
and $E_{{\bf k}}=\xi_{{\bf k}}$. Within the present partial
charge-spin separation fermion-spin framework, the dressed holons
and spinons move self-consistently, where $T^{(a)}_{c}$ and other
order parameters are determined by the self-consistent equations
(12) and (14) in the condition $\Delta^{(a)}_{h}=0$. For small
dopings, the dressed holons are concentrated around the wave
vector ${\bf k}\sim (0,0)$ , then from Eq. (13b) we find that
$T^{(a)}_{c}\propto \rho^{(a)-1}(0)\delta$, with $\rho^{(a)}(0)$ is
the dressed holon density of state at the corresponding chemical
potential $\mu$. Although the SC state is characterized by the
charge-spin recombination, the main physical properties of the SC
state are dominated by charged holons. This is why the superfluid
density in the underdoped regime vanishes more or less linearly
with concentration of doped holes, and the doped cuprates are the
hole-type gossamer superconductors.

\begin{figure}[prb]
\epsfxsize=3.5in\centerline{\epsffile{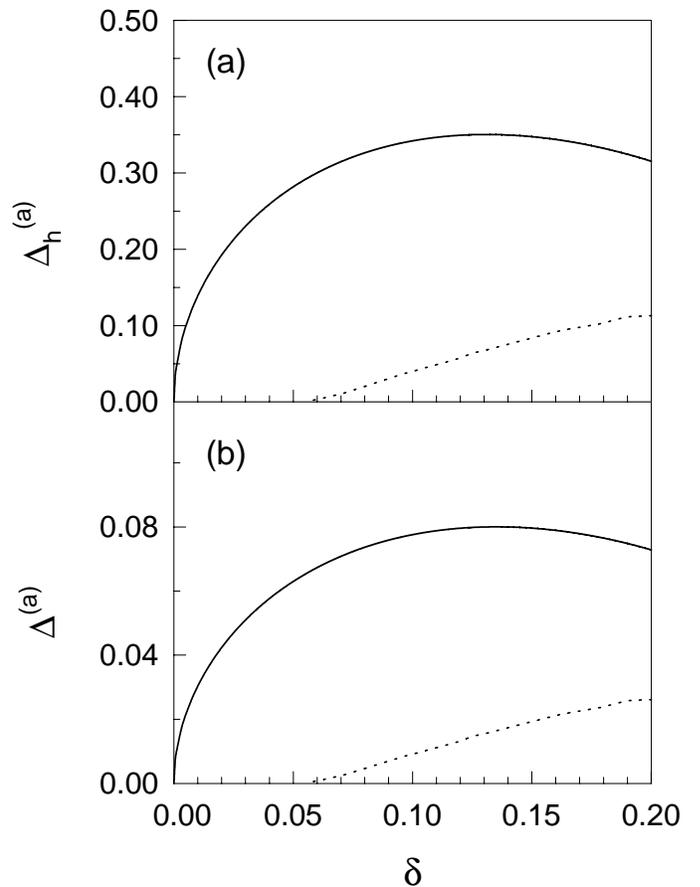}}
\caption{The dressed holon (a) and superconducting (b) gap
parameters in the s-wave symmetry (solid line) and d-wave symmetry
(dashed line) as a function of hole doping concentration at
$T=0.005J$ and $t/J=2.5$.}
\end{figure}
\begin{figure}[prb]
\epsfxsize=3.5in\centerline{\epsffile{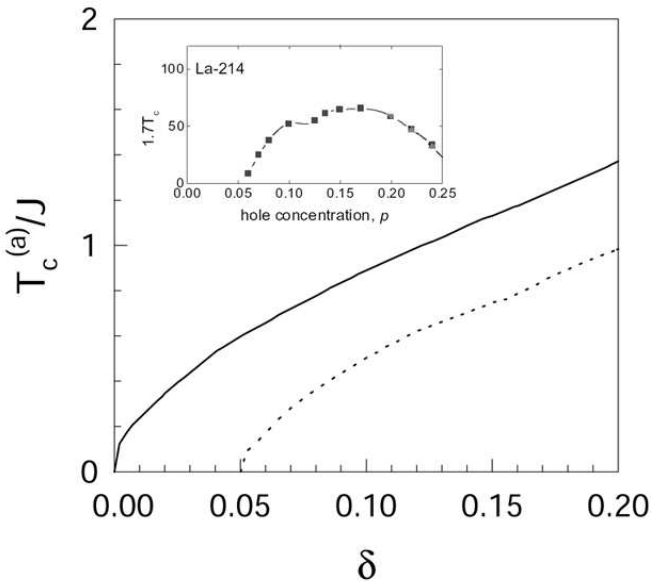}}
\caption{The superconducting transition temperature as a function
of hole doping concentration in the s-wave symmetry (solid line)
and d-wave symmetry (dashed line) for $t/J=2.5$. Inset: the
experimental result taken from Ref. [5].}
\end{figure}

The attractive interaction between dressed holons in Eq. (10b) is
induced by exchanging dressed spinon excitations, and therefore is
determined by the weight of the dressed spinon spectrum function.
In the above calculation of the effective dressed holon (then SC)
gap function (10b), we have replaced the full dressed spinon
Green's function $D(p)$ by the MF dressed spinon Green's function
$D^{(0)}(p)$, this leads to favor the s-wave pairing state,
$\Delta^{(s)}_{h}({\bf k})\mid_{{\bf k}_{c}}\sim\Delta^{(s)}_{h}$,
and the d-wave pairing state is suppressed, $\Delta^{(d)}_{h}
({\bf k})\mid_{{\bf k}_{c}}\sim 0$, since in the MF level the main
contribution for the weight of the dressed spinon spectrum
function comes from dressed spinon excitations around wave-vector
${\bf k}_{c}\sim [\pi,\pi]$. However, the s-wave pairing state is
suppressed, and the d-wave pairing state is enhanced if the full
dressed spinon Green's function $D(p)$ is considered. Since in this
case, the dressed spinon self-energy renormalization due to the
dressed holon-spinon interaction leads to the incommensurate spin
fluctuation \cite{feng1,feng6}, then the main contribution for
the weight of the dressed spinon spectrum function comes from the
renormalized dressed spinon excitations around wave-vectors
${\bf k}_{c}\sim [(1\pm \varsigma)\pi,\pi]$ and ${\bf k}_{c}\sim
[\pi,(1\pm \varsigma)\pi]$, which is favorable for the d-wave
pairing state, $\Delta^{(d)}_{h}({\bf k})\mid_{{\bf k}_{c}}\sim
\Delta^{(d)}_{h}[1+{\rm cos}(\varsigma\pi)]$, and not for the
s-wave pairing state, $\Delta^{(s)}_{h}({\bf k})\mid_{{\bf k}_{c}}
\sim \Delta^{(s)}_{h}[1-{\rm cos}(\varsigma\pi)]$. In fact, the
d-wave gap function
$\Delta^{(d)}({\bf k})\propto {\rm k}^{2}_{x}-{\rm k}^{2}_{y}$
belongs to the same representation $\Gamma_{1}$ of the orthorhombic
crystal group as does the s-wave gap function
$\Delta^{(s)}({\bf k})\propto {\rm k}^{2}_{x}+{\rm k}^{2}_{y}$,
the two perhaps can mix at will, and there is some evidence from
experiments to support this symmetrical picture \cite{brawner}.
This is also why some experiments \cite{jacobs} can be fitted well
by both s-wave and d-wave symmetries.

In the normal-state, the electron diagonal Green's function
$G(i-j,t-t')=\langle\langle C_{i\sigma}(t);C^{\dagger}_{j\sigma}
(t')\rangle\rangle$, which is a convolution of the spinon Green's
function $D(p)$ and diagonal holon Green's function $g(k)$, has
been calculated within the fermion-spin theory \cite{feng2}. With
the help of this electron diagonal Green's function, the physical
properties of the electron spectral function $A(k)=-2{\rm Im}G(k)$
have been discussed \cite{feng2,feng8}, and the results are
qualitatively consistent with ARPES experiments \cite{shen5}. This
electron spectral function has been used to extract the electron
momentum distribution (then the electron Fermi surface) \cite{yuan}
$n_{{\bf k}}=\int^{\infty}_{-\infty}d\omega A({\bf k},\omega)n_{F}
(\omega)/2\pi$, and the results show that $n_{{\bf k}}$ for the
2D $t$-$J$ model does not follow the behavior expected from
Luttinger's theorem, in agreement with the numerical results
\cite{putikka}. The holons center around $[0,0]$ in the Brillouin
zone, then the charge-spin recombination from the convolution of
the spinon Green's function and holon Green's function leads to
form the electron Fermi surface, therefore the electron is a
composite particle, and could account for the spread of low energy
excitations throughout the Brillouin zone
\cite{anderson2,putikka,yuan}, these are consistent with the
results found in ARPES experiments \cite{shen5}. The SC
fluctuations at low temperatures in the 2D $t$-$J$ model from a
higher temperature state which can not be described as a Fermi
liquid \cite{anderson1}. Since the electron Cooper pairing state
is originated from the holon pairing state as mentioned above,
then the electron gap is induced by the holon gap. In this case,
both holon and electron gaps are responsible for the SC state.

Finally, we have noted that an obvious weakness of the present
theoretical results is that $T^{(a)}_{c}$ is too high, and not
suppressed in the overdoped regime. Recently, ARPES measurements
\cite{ding} have shown that $T_{c}$ in doped cuprates is
dependent on both the gap parameter and weight of the coherent
excitations in the spectral function $Z_{A}$, while this
$Z_{A}<1$ is closely related to the antisymmetric part of the
self-energy function $\Sigma^{(h)}_{1}(k)$, and increases
monotonically with increasing dopings in the underdoed and
optimally doped regimes, and then slows down in the overdoped
regime. This experimental result strongly suggests that the single
particle coherence plays an important role in HTSC. In this case,
it is then possible that the weakness perhaps due to dropping
$\Sigma^{(h)}_{1}(k)$ in Eq. (10a) in the present case may be
cured by considering this self-energy function, and these and
other related issues are under investigation now.

In summary, we have discussed the mechanism of HTSC in doped
cuprates based on the partial charge-spin separation fermion-spin
theory. Within the $t$-$J$ model, it is shown that dressed holons
interact occurring directly through the kinetic energy by
exchanging dressed spinon excitations, leading to a net attractive
force, then the electron Cooper pairs originating from the dressed
holon pairing state are due to the charge-spin recombination, and
their condensation reveals the SC ground state. The SC transition
temperature $T^{(a)}_{c}$ is determined by the dressed holon pair
transition temperature, and is proportional to the concentration of
doped holes in the underdoped regime. With the common form of the
electron Cooper pair, we also show that there is a coexistence of
the electron Cooper pair and short-range AF correlation, and hence
the short-range AF fluctuation can persist into the SC state.
These results are qualitatively consistent with experiments
\cite{uemura,tallon,kastner} in the underdoped regime. In the
present picture, each lattice site is singly occupied by a spin-up
or -down electron at the half-filling, then the spins are coupled
antiferromagnetically without AFLRO. With dopings, dressed holons
move in the dressed spinon liquid background, and form pairs by
exchanging dressed spinon excitations at low temperature, then
these dressed holon (then electron) pairs condense to the SC state,
which is not far in spirit from the original resonant valence bond
theory \cite{anderson3}.

\acknowledgments
The author would like to thank Dr. Ying Liang, Dr. Jihong Qin,
and Dr. Tianxing Ma for the helpful discussions. This work was
supported by the National Natural Science Foundation of China
under Grant Nos. 10125415 and 10074007.

\end{document}